# BACKWARD-EMITTED SUB-DOPPLER FLUORESCENCE FROM AN OPTICALLY THICK ATOMIC VAPOR

João Carlos de Aquino Carvalho[1], Athanasios Laliotis[1],
Martine Chevrollier[2], Marcos Oriá[2], Daniel Bloch[1]

[1]Laboratoire de Physique des Lasers, UMR 7538 du CNRS, Université Paris13 - Sorbonne Paris Cité, 93430 Villetaneuse, France

[2]Universidade Federal Rural de Pernambuco, UACSA, Cabo de Santo Agostinho, PE, Brazil

*Corresponding author: daniel.bloch@univ-paris13.fr

*Literature mentions only incidentally a sub-Doppler contribution in the excitation spectrum of the backward fluorescence of a dense vapor. This contribution is here investigated on Cs vapor, both on the first resonance line (894 nm) and on the weaker second resonance line (459nm). We show that in a strongly absorbing medium, the quenching of excited atoms moving towards a window irradiated under near normal incidence reduces the fluorescence on the red side of the excitation spectrum. Atoms moving slowly towards the window produce a sub-Doppler velocity-selective contribution, whose visibility is here improved by applying a frequency-modulation (FM) technique. This sub-Doppler feature, induced by a surface quenching combined with a short absorption length for the incident irradiation, exhibits close analogies with the narrow spectra appearing with thin vapor cells. We also show that a normal incidence irradiation is essential for the sub-Doppler feature to be observed, while it should be independent of the detection geometry.*

Fluorescence detection in a laser-excited gas cell provides a variety of useful information. In particular, it both evaluates the absorption processes when sweeping the frequency of the exciting laser (excitation spectroscopy) and diagnoses decay mechanisms —including energy level transfers— when the detected fluorescence is wavelength-selected. In the course of our detailed studies of the non-zero temperature quantum effect on the atom-surface interaction for highly excited atoms [1], we investigated the fluorescence of a dense Cs vapor close to a strongly heated window, in order to evaluate locally the vapor density, atomic velocity, and rate of collisions or radiative transfer. The fluorescence observed from the entrance window of the cell is a natural tool for this purpose.

Standard high-resolution atomic spectroscopy is hard to apply in an optically thick medium because of various broadenings, mainly due to the high atomic density associated to optical thickness (yielding collisions and velocity redistribution), and to propagation, with its strong absorption possibly accompanied by radiation-trapping. This is why literature has addressed sub-Doppler spectroscopy in an optically thick medium only infrequently [2-7]. Rather, specific single beam techniques based upon a confinement of the detection region for high densities, such as reflection spectroscopy (selective reflection or evanescent wave spectroscopy), or thin vapor cell, have been shown to provide sub-Doppler signals (see [8] for a review). Remarkably, in a series of papers [3-6] on so-called "retrofluorescence" —*i.e.* observation of the fluorescence in a backward direction, relatively to an excitation under near normal incidence— the J.-M. Gagné group has observed [3,4] a tiny sub-Doppler component (width ~40 MHz *vs.* Doppler width ~250 MHz) in the excitation spectrum of fluorescence on the Cs $D_2$ line at 852 nm appearing over a broad dip in the fluorescence spectrum (see also in [7] for a confirmation on the Rb $D_2$ line, with a width ~25-30 MHz for the sub-Doppler contribution).

The broad spectral dip [9,10] is known to originate in a non-radiative relaxation on the wall (the corresponding quenching is mostly on resonance, corresponding to maximal absorption length), but the specific sub-Doppler feature reported in [3,4,7] has lacked a clear interpretation. The claim has been that it is needed to discriminate between a far-field region (*i.e.* standard free vapor) and a "vapor boundary layer" or "near-field proximity region", whose thickness would be ~$\lambda$ (the optical wavelength), and where a thin metal layer —possibly sub-mono-layer— coating the window would dramatically influence the atomic lifetime [3]. At the opposite, from our own experience in a variety of spectroscopy experiments close to an interface (see [1, 8, 11-13] and refs. therein), spectroscopy in the vicinity of a window can be fully interpreted by standard processes of atomic physics once the interaction with light is described in the transient regime, as long as the sudden de-excitation on the window or the transient coupling to the irradiation for atoms leaving the window are taken into account. Moreover, in our present work, the window of one of our vapor cells [1] is largely overheated relatively to the colder Cs reservoir, preventing any metallic condensation (residual Cs atoms or clusters would be removed by thermal desorption).

Here, we report on fluorescence experiments in an optically thick regime, performed for two different transitions of Cs, namely the $D_1$ resonance line ($6S_{1/2}$-$6P_{1/2}$ at 894 nm), and the weaker (~10 times) second resonance line $6S_{1/2}$-$7P_{1/2}$ at 459 nm (fig. 1). For these lines, the hyperfine components are limited to F={3,4}→F'={3,4}, avoiding a hyperfine structure as large as the Doppler broadening, as it is the case on the Cs (or Rb) $D_2$ line. Our results confirm the presence of a sub-Doppler contribution even on a weak line and at moderate densities, as easily evidenced through a frequency modulation (FM) technique. We show also that the dip [2, 9-10] around the line center is actually red shifted, confirming that the wall relaxation applies mainly to atoms moving towards the window. As for the many other sub-Doppler features [8, 11-12] appearing for an excitation under normal incidence in the vicinity of an interface, we show that the sub-Doppler contribution is associated to a specific response of the atoms both in the vicinity of the surface and whose motion —counted along the normal to the window— is slow.

A schematic of the experimental set-up is provided in fig.2, and the experiments were conducted for an irradiation under near normal incidence —up to 5° for the 894 nm set-up, 1-2° for the 459 nm experiment.

Due to the cell and oven geometry, the backward-emitted fluorescence was detected slightly off-axis, under an angle up to 10°, on the side opposite to the one of the irradiation in order to better discriminate between fluorescence and scattered laser light. Although the issue was not addressed in the previous literature, our interpretation will show that the detection angle should not affect the sub-Doppler contribution of interest. The fluorescence light was focused onto a photodetector collecting a solid angle ~ $10^{-2}$ sr. The experiments were performed on a standard Cs glass cell (1 cm-long) for the 894 nm experiment. For the 459 nm experiment, we used the ~ 8 cm-long Cs cell described in [1], whose sapphire window is superpolished and annealed. In this last case, a three-section oven allows a major overheating of the cell relatively to the Cs reservoir (here, the window has been kept at ~300°C, for a Cs reservoir temperature below 200 °C). The narrow linewidth (up to a few MHz) laser sources were a DBR laser diode for 894 nm, and an extended cavity laser diode for 459 nm line. For both of them, a frequency modulation (FM) could be applied. The irradiating power falling onto the cell was kept low enough (< 1 mW/mm²) to avoid strong saturation —which in any case would decrease quickly with propagation and absorption. For all experiments, an auxiliary saturated absorption (SA) spectrum was recorded simultaneously to provide sub-Doppler references of the atomic transitions. For the 894 nm line, the fluorescence is essentially emitted on the excitation line itself at 894 nm, although energy-pooling collisions may allow fluorescence from highly excited levels, including the $D_2$ line at 852 nm. The 459 nm line brings the atom in a high-lying state, and numerous excited atomic levels can be reached by fluorescence decay or by collisions, which are even susceptible to populate energy levels higher than 7P (see fig.1). We have been able to observe such excited populations through the analysis of the emitted fluorescence, but for the Cs densities used for the present report, their influence should remain extremeùy marginal. Here, the optical thickness also allows radiation trapping to occur, mostly on the resonance doublet through cascades. To detect the fluorescence induced by the 459 nm absorption and to eliminate the scattering of the incident light, we use a Ge photodetector, sensitive up to 1.7 μm, after a filter which blocks wavelengths shorter than ~ 1 µm. Hence, we detect (see fig.1) the emission at ~ 1.38 µm (from the 7 $P_{1/2}$ level to $5D_{3/2}$), and the 1.36 µm and 1.47 µm doublet from $7S_{1/2} \rightarrow 6P$.

In the optically thin regime, the excitation spectrum of fluorescence is simply a Doppler-broadened resonance whose amplitude increases with density; when the Cs density is increased and becomes so large that the laser light is totally absorbed in the volume allowed by the field of view of the detector (typically, for our set-up, the fluorescence is collected on a depth hardly exceeding a few mm), the fluorescence intensity reaches a limit while the spectrum broadens. This maximal intensity is simply the equivalent, with respect to the solid angle of detection, of the incident light intensity which has been totally converted into fluorescence. Although radiation trapping can induce multiple cycles of absorption and emission, the flux of emerging light does not increase, as governed by the incident flux of photons.

For the 894 nm line, the fluorescence signal starts to saturate for a density ~ $5.10^{13}$ at.cm$^{-3}$ (*i.e.* 120 °C). The two hyperfine components (~ 1.1 GHz splitting for the excited state $6P_{1/2}$), naturally resolved with respect to the Doppler broadening (~ 200 MHz) at low Cs density, are still resolved in the conditions of an optically thick medium, but, as shown on top of fig.3, the maximal fluorescence is attained for a range of frequencies exceeding the Doppler width, corresponding to total absorption through the cell. In this regime of optically thick medium, the plateau drops down only when the frequency detuning is sufficient for the overall absorption to decrease, so that at high Cs densities, the wings of the plateau are located well apart from the Doppler-broadened central frequency. The most remarkable feature is that a dip can be seen on this plateau of maximal intensity. Such an effect has been previously noticed (see [9] and also [3-5]) and results from the fact that when the absorption occurs close to the window (*i.e.* for a frequency close to line center), surface collisions can interrupt non radiatively the (multiple) cycle(s) of fluorescence: the dip is hence the signature of this frustrated fluorescence. These features are similar to those reported previously for excitation on the Cs 852 nm line [3], but here the resolution of the hyperfine components makes visible that the dip is shifted to the red (note that the discussion in [3,4,7] has been limited to the strongest hyperfine component, for which the red side is plagued with the —hardly resolved— other hyperfine components, of a weaker amplitude). It is however very delicate to find a narrow contribution on the fluorescence spectrum itself, as it was already the case in [3,4,7]. However, the FM applied to the laser (at 10 kHz, and with an excursion ~ 10 MHz much smaller than the width of the direct signal) allows to process the signal by a lock-in detector, delivering the frequency-derivative of the initial spectrum, which emphasizes the narrow contributions. For a large range of densities, the FM detection makes visible sub-Doppler contributions (see fig. 3), which appear on the red side of the auxiliary SA marker of each component in the dip of fluorescence. The SA marker is naturally associated to atoms with a null (longitudinal) velocity in the lab frame, whose resonance is not shifted by the Doppler shift.

Very similar observations are performed on the line at 459 nm (fig. 4), which is much weaker and nearly insensitive to radiation trapping, as long as the Cs density is increased in order to reach an absorption comparable to the one for the $D_1$ line measurements. The signal-to-noise ratio benefits from the wavelength filtering which eliminates the scattering of the incident irradiation. However, the $7P_{1/2}$ hyperfine structure (377 MHz) remains unresolved, owing to the Doppler broadening, twice as large as for the 894 nm line. Here too, the excitation lineshape of fluorescence broadens when the atomic density is high enough to make the detected volume an optically thick region. On the direct signal, a plateau and a dip also appear, with a clear asymmetry. The dip is notably shifted to the red, distinctly on the red side of the marker of the F=4→F'=3 hyperfine component. No significant differences are found between the two hyperfine doublets F=3→{F'=3, 4} (the strength of the F=3→F'=4 component is 3 times larger than the F=3→F'=3), and F=4→{F'=3, 4} (the two components of the doublet F=4→F'=3 and F=4→F'=4 are of a comparable strength, in a ratio of 5 to 7). Through the FM detection (modulation at ~1 kHz, excursion ≤ 10 MHz), narrow sub-Doppler features are made observable. They also appear on the red side —with respect to each hyperfine SA marker— and vary only moderately with the Cs density. All these features for the excitation on the 459 nm line confirm the strong analogy with the observations on the 894 nm line.

To interpret what appears as a general behavior, with a sub-Doppler dip on the red side of the spectrum in addition to the previously-known broad dip, we have to consider in detail the fluorescence reduction occurring specifically in the vicinity of the window. Reduction of the fluorescence efficiency can occur either in a direct process of excitation and fluorescence —hence restricted to atoms moving *towards* the surface, with their red-shifted excitation resonance— or during a more complex coupling between the initial excitation and the detection of the fluorescence, involving a transfer from one atom to another (*e.g.* by radiation trapping or collision) or at least a velocity change between excitation and fluorescence (*e.g.* velocity changing "weak" collision). In this last case, the surface quenching of an atom moving towards the surface occurs whatever is the laser frequency of the initial excitation process: the redistribution of excitation may transport an excitation initially induced by a blue-shifted irradiation —*i.e.* addressing atoms flying away from the surface— to an atom moving towards the window and sensitive to quenching, as it was analyzed in [10] (although for densities much higher than ours). Here, the typical thermal velocity for Cs is ~200-250 m/s while the excited state lifetime is ~30 ns for Cs (6P), and ~ 100 ns for Cs(7P). This shows that the atoms the most susceptible to undergo a wall-induced relaxation, instead of a sudden fluorescent emission at a random time, are located in the vicinity (~10-30 μm) of the window. This compares with the absorption length (on resonance) at the atomic densities of interest, and justifies to model simply the detected fluorescence as proportional to $I_{\text{fluo}}$, with:

$$I_{\text{fluo}} = \int_{-\infty}^{+\infty} \frac{dv\, f(v)}{(\gamma/2)^2 + (\delta - kv)^2} \int_0^{+\infty} dz\, \exp(-z/\Lambda)\, \eta(z,v) \quad (1)$$

In eq.(1), the first integral, over the velocity distribution $f(v)$ (one-dimensional distribution of velocity —along the normal to the window— and assumed to be Maxwellian) describes a simple excitation governed by a Doppler-shifted Lorentzian response to the irradiation (where $\delta$ is the detuning from resonance, and $k$ the wavevector modulus). The second integral is for a spatial integration (along the normal to the window) and assumes the excitation to be proportional to the local irradiation intensity following an exponential decay characterized by an absorption length $\Lambda$, while the conversion of the atomic excitation into a detectable fluorescence is governed by an efficiency factor $\eta(z, v)$, which is velocity-dependent because of the quenching by the window. In a simple case (*i.e.* neglecting redistributions), one assumes $\eta(z, v) = [1- \exp(z/v\tau)]$ for $v<0$ and $\eta(z,v) =1$ for $v \geq 0$, with $\tau$ the lifetime of the relevant excited level (*i.e.* decay time for the conversion of excitation into fluorescence) and $v <0$ corresponding to atoms directed towards the window. Note that among the simplifying assumptions of eq.(1), the atomic excitation obeys a Lorentzian response because the effects of the transient regime of interaction in the building-up of the excited population have been neglected. Although these transient effects may become important [8, 14] in the vicinity of a surface, they would reduce the efficiency of excitation only for fast departing atoms, or in the very close vicinity of the surface when the surface interaction modifies the energetic structure of the atom.

Figures 5-7 illustrate the predictions derived from eq.(1), showing that the velocity $\Lambda/\tau$ is the key parameter to discriminate atoms sensitive to the quenching, as can be expected from the definition of $\eta(z, v)$,. Fig.5 shows the decrease of the fluorescence signal on the red side of the atomic resonance, with respect to the fluorescence spectrum in the absence of quenching –*i.e.* the Gaussian-shaped fluorescence for $\eta(z,v)=1$—, as long as $\Lambda/\tau$ is only a fraction of the thermal velocity u (only atoms with velocity such as $-\Lambda/\tau < v<0$ are insensitive to the quenching). Figure 6 addresses a more realistic situation for the optical thickness of the medium, thick only on-resonance, while returning to quasi-transparency for a detuned irradiation frequency. The additional assumptions of an absorption length $\Lambda$ varying according to a usual Voigt profile of absorption –resulting in a Beer-Lambert propagation—, and of a physically bounded spatial integration of the fluorescence emitters, allow to predict a fluorescence excitation

spectrum combining the known flat response spanning well over the Doppler width, with a significant drop of the fluorescence on a red part of the spectrum owing to the velocity-selective quenching in eq. (1) through $\eta(z,v)$ which affect atoms going towards the window. The predictions are clearly reminiscent of the observed spectra, as presented in figs. 3-4. Varying the parameters in a way mimicking the change of Cs vapor density with the cell temperature does not affect the essence of the global spectral shape. The FM counterpart of fig.6 is provided in fig.7 (with $dI_{fluo}/d\delta$), showing again clear analogies with the experimental spectra for both the 894 nm and the 459 nm lines.

The simplified eq. (1) is successful enough to establish that the essence of the sub-Doppler response is the combination of a velocity-dependent loss term $[1-\eta(z,v)]$, favoring the response of slow atoms, and of the short effective length $\Lambda$ of the excitation region in the optically-thick vapor, which allows sensitivity to the surface quenching. A fully quantitative analysis, aiming notably to predict the relative amplitude of the narrow contribution with respect to the plateau of maximal fluorescence, would be an overwhelming task. This is because of a variety of processes susceptible to affect eq.(1), and because at the high atomic densities associated to optically thick media, there is a number of complex processes (multilevel transfer, energy-pooling collisions) which scatter the initial excitation. As a first limit of a quantitative applicability of eq.(1), we have already mentioned that the surface quenching can also reduce the fluorescence signal for an irradiation on the blue-side, when the initial excitation to atoms moving away from the surface is transferred to atoms moving towards the surface: because of this, the dip observed in the fluorescence spectrum may appear shifted to the red, without being limited to the red side of the excitation spectrum, before getting totally symmetrized. Another difficulty arises for evaluating the absorption length, or the spatially-dependent irradiation $exp(-z/\Lambda)$: for a fluorescence emitted at the same wavelength as the one of the irradiation (*e.g.* 894 nm line), the emitted fluorescence is partly reabsorbed, hence reducing the effective length of propagation $\Lambda$ in the overall process detected through fluorescence; moreover, for a rather high intensity at the entrance of the window —a situation quite natural because the intensity, despite saturation, quickly attenuates through propagation— the local irradiation no longer follows an exponential law. The frequency-dependent propagation of the irradiation itself may become very complex (see [15] for an extreme case), notwithstanding the already mentioned transient effects affecting the atomic excitation and the absorption itself. The efficiency of the quenching, governed in eq. (1) by $[1-\eta(z,v)]$ and hence by the decay time $\tau$ of the fluorescence, is also uneasy to determine in the presence of a transfer of excitation partly preserving the initial velocity $v$ (e.g. fluorescence cascade or even velocity-changing collisions), as the effective path to the window may become longer than a free flight. In addition, the atom radiative lifetime itself is susceptible to be space-dependent, through a possible quenching into a surface mode [16], or by the weaker coupling to evanescent modes [17] which is a general phenomenon.

The successful interpretation by eq. (1) allows to understand the geometry needed for the observation of a narrow contribution, which was never discussed in previous reports of a sub-Doppler structure [3-5,7]. The velocity-dependent loss term is intrinsically dependent on the velocity normal to the window, while for an irradiation under an oblique incidence, the Lorentzian absorption factor would be sensitive to a Doppler shift governed by the relevant oblique component of velocity, finally yielding a residual Doppler broadening to the sub-Doppler structure governed by $\Lambda/\tau u$. Conversely, if the fluorescence is collected under a large angle, or under an oblique emergence, there is no need to modify the velocity-dependent loss term (eventually, the effective attenuation length $\Lambda$ may change when the fluorescence is detected on the same line —strongly absorbed— as the incident irradiation). Our prediction of an insensitivity of the sub-Doppler response to the orientation of the detection is clearly in agreement with the early observation [18] of a sub-Doppler feature in the excitation of the fluorescence emitted in an evanescent field, which is nothing else than an extremely large detection angle.

The present observation of a narrow contribution originating in the selected response of atoms with a slow (normal) velocity is connected to the general class of sub-Doppler signatures [8, 11-12], when a single normal beam is sent to a surface or a thin cell and where the contribution of the slowest atoms — along the normal— becomes dominant because of a long interaction time, which supersedes the deleterious effect of the transient interaction. The analogy is particularly close with the fluorescence spectrum in very thin vapor cells [11], where the excitation spectrum (under normal incidence) is sub-Doppler because only the slowest atoms contribute efficiently to the fluorescence. Here, the strong optical attenuation is similar to the abrupt interruption of the fluorescence by the windows of a micrometric thin cell. The difference is that in our present case, the strong attenuation on a short distance is a "soft" exit window, not imposing a quenching, so that the sub-Doppler contribution appears only on the red-side of the transition (instead of being intrinsically symmetric for a thin cell). To make the analogy more obvious, in one of the curves of fig. 5, the velocity-dependent loss term in eq. (1) has been replaced by a symmetric loss term, namely with $\eta(z,v) = \eta(z,-v)$ for $v > 0$. This makes the overall spectrum symmetric around the resonance, while the response on the red-side is practically unaffected. This response with the symmetric loss term, clearly sub-Doppler as long as $\Lambda/\pi u \ll 1$, exhibits clear analogies with the experimental observations in thin cells [11]. This confirms again our analysis that the sub-Doppler structure is insensitive to the detection angle as the experimental observations of sub-Doppler fluorescence in thin cells have been usually performed at ~ 90° from the normal to the window [11].

To summarize, our experiments confirm the general possibility of observing, in a medium strongly absorbing in the vicinity of the window, a narrow sub-Doppler signature in the excitation spectrum of fluorescence, made more visible when implementing a FM technique. This narrow contribution has no connection with the presence of any metallic film [3-7]. Rather, our interpretation evidences the analogy with various sub-Doppler responses observed with a single laser, and occurring in the vicinity of a window. This justifies why a near normal incidence is required for the excitation, while the fluorescence can be detected in any direction.

Finally, we have found (on the 459 nm line) that the sub-Doppler feature is robust when a wavelength-selection is applied to the detection scheme, and this allows the fluorescence to be detected on a zero-background, insensitive to the non resonant scattering by the window. Further discrimination on the direct or cascade fluorescence (*e.g.* from 7P, or from 7S) may provide complementary information owing to the different time constants involved. Although the narrow contribution may remain of a small amplitude because a strong absorption required usually comes with a collisional broadening, the sub-Doppler feature may help to analyze how some excited or rare species survive in the vicinity of the window, where specific charges or stray fields may be present. Also, it can be expected that the sub-Doppler signature is not present for a detection from a level populated solely by collision-induced transfers (and velocity redistribution). Conversely, the sub-Doppler structure should survive when the population is induced by internal transfers (fluorescence cascade, blackbody transfer [19]...), or when strong absorption can be associated to another species (*e.g.* dimers *vs.* monomers [20], rare isotope in a mixture); in a similar way, diagnostics of the behavior of excited states close to a surface may be operated in a dual laser excitation scheme, when the medium is optically thick only for the initial pumping step.

**Acknowledgment.** Franco-Brazilian cooperation was supported by CAPES-COFECUB Ph740/12. J.C.A.C. acknowledges financial support for his PhD preparation by "Ciências sem fronteiras" (Brazil).

**References**

[1] A. Laliotis, T. Passerat de Silans, I. Maurin, M. Ducloy, D. Bloch, "Casimir-Polder interactions in the presence of thermally excited surface modes," Nature Comm. **5**, 5634 (2014).

[2] S. Svanberg, G.-Y. Yan, T. P. Duffey, A. L. Schawlow, "High-contrast Doppler-free transmission spectroscopy", Opt. Lett. **11**, 138 (1986)

3] K. Le Bris, J.-M. Gagné, F. Babin, M.-C. Gagné, "Characterization of the retrofluorescence inhibition at the interface between glass and optically thick Cs vapor," JOSA B **18**, 1701 (2001).

[4] J.-M. Gagné, C. K. Assi, K. Le Bris, "Measurement of the atomic Cs [$6^2P_{3/2}$ ($F_e$= 5)] hyperfine level effective decay rate near a metallic film with diode laser retrofluorescence spectroscopy," JOSA B **22**, 2242 (2005)

[5] K. Le Bris, C. K. Assi, J.-M. Gagné, "Spectroscopic investigation of sensitized-Cs ($6^2P_{3/2}$ -$6^2P_{1/2}$ ) laser retrofluorescence in a pure optically thick vapor near a dissipative surface" Can. J. Phys. **82**, 387 (2004)

[6] J.-M. Gagné, K. Le Bris, M.-C. Gagné, "Laser energy-pooling processes in an optically thick Cs vapor near a dissipative surface" JOSA B **19**, 2852 (2002)

[7] Jing Liu, Yifan Shen, Kang Dai, "Measurement of the $^{87}$Rb[$5P_{3/2}$ (F=3)−$5S_{1/2}$(F=2)] Effective Nonradiative Relaxation Rate near a Metallic Film", International Journal of Spectroscopy, doi:10.1155/2010/704183, Article ID 704183 (2010)

[8] D. Bloch and M. Ducloy, "*Atom-Wall Interaction*", Adv At. Mol. Opt. Phy., vol.50, B. Bederson and H. Walther eds., Elsevier Academic Press, San Diego, 2005 pp. 91- 156

[9] G Zajonc, and AV Phelps, "Nonradaitive transport of atomic excitation in Na vapor" Phys. Rev. A **23**, 2479 (1981)

[10] H. van Kampen, V. A. Sautenkov, A. M. Shalagin, E. R. Eliel, J. P. Woerdman, Dipole-dipole collision-induced transport of resonance excitation in a high-density atomic vapor" Phys. Rev. A **56**, 3569 (1997)

[11] D. Sarkisyan, T. Varzhapetyan, A. Sarkisyan, Yu. Malakyan, A. Papoyan, A. Lezama, D. Bloch, M. Ducloy, "Spectroscopy in an extremely thin vapor cell: Comparing the cell-length dependence in fluorescence and in absorption techniques", Phys. Rev. A **69**, 065802 (2004)

[12] S. Briaudeau, D. Bloch, M. Ducloy, "Detection of slow atoms in laser spectroscopy of a thin vapor film", Europhys.Lett. **35**, 337 (1996).

[13] P. Ballin, E., Moufarej, I. Maurin, A. Laliotis, D. Bloch , "Three-dimensional confinement of vapor in nanostructures for sub-Doppler optical resolution", App. Phys. Lett., **102**, 231115 (2013).

[14] Even for a two-level system, fluorescence from an excited population is at least a second order process, intrinsically more complex process than the coherent first order linear effect responsible for selective reflection or absorption in a microcell (see *e.g.* [8,11,12]). This explains why, until now, a rigorous theory of sub-Doppler fluorescence in a thin vapor cell has been limited to a formal approach - see: G. V. Nikogosyan, D. G. Sarkisyan, and Yu. P. Malakyan, "Absorption of resonance radiation and fluorescence of a layer of an atomic gas with thickness of the order of a wavelength ", J. Opt. Technol. **71**, 602 (2004).

[15] L. Roso-Franco, "Propagation of light in a nonlinear absorber", JOSA B **4**, 1878 (1987)

[16] H. Failache, S. Saltiel, A. Fischer, D. Bloch, M. Ducloy, "Resonant Quenching of Gas-Phase Cs Atoms Induced by Surface Polaritons", Phys Rev. Lett. **88**,243603 (2002)

[17] W. Lukosz and R.E. Kunz, "Fluorescence lifetime of magnetic and electric dipoles near a dielectric interface", Opt. Comm., **20**, 195 (1977)

[18] A.L.J. Burgmans, M.F.H. Schuurmans, B. Bölger, "Transient behavior of optically excited vapor atoms near a solid interface as observed in evanescent wave emission," Phys. Rev. A, **16**, 2002 (1977).

[19] M. Pimbert "Transfert d'excitation électronique, par collision atomique, entre niveaux élevés d'un atome de césium" J. Phys. (Paris), **33**, 331 (1972)

[20] T.Passerat de Silans, I. Maurin, A. Laliotis, P.Chaves de Souza Segundo, D. Bloch, "Extra sub-Doppler lines in the vicinity of the third resonance 6S-8P of atomic Cs attributed to optically induced Cs dimers", Phys. Rev. A **83**, 043402 (2011)

**Figure Captions**

Fig.1 : Scheme of relevant Cs levels, showing the first and second resonance lines (894 nm, and 459 nm respectively), with the most relevant transitions for fluorescence. The 7P level decays to 7S (with further cascades in the IR to 6P, detectable in our set-up), or to 5D ( detectable IR fluorescence), or along the second resonance itself at 459 nm, with respective branching ratios ~50%, ~20%, ~30 %.

Fig. 2: Schematic of the experimental set-up. The nearly resonant laser is sent onto the vapor cell close to the normal incidence, here shown as a dashed line, and is rapidly absorbed by the dense vapor. The emitted fluorescence is focused onto the photodetector (solid angle ~$10^{-2}$ sr) and possibly filtered in wavelength. The signal $I_{fluo}$ from the photodetector can be processed with a lock-in amplifier (for "FM fluorescence").

Fig. 3: Excitation frequency spectrum of the backward emitted fluorescence on the Cs $D_1$ line (894 nm) around the F=4 →F' ={3,4} hyperfine components (splitting: 1.17 GHz): on top is the direct fluorescence for Cs reservoir at 130 °C ( ~ $8.10^{13}$ at.cm$^{-3}$), the middle is the corresponding simultaneous recording of the FM signal; the bottom is the zoomed FM signal, with the 2 additional Cs temperatures 120°C (~ $5.10^{13}$ at.cm$^{-3}$) and 150°C (~ $2.10^{14}$ at.cm$^{-3}$). The experiment is performed on a Cs cell made of glass, the temperature of the window is ~ 200 °C, and the frequency markers are provided by the corresponding auxiliary SA reference.

Fig. 4: Excitation frequency spectrum of the backward emitted fluorescence (filtered between 1.1 and 1.7 µm) on the Cs 459 nm around the F=4 →F' ={3,4} hyperfine components (splitting: 377 MHz): on top is the direct fluorescence for Cs reservoir at 170 °C, the middle is the corresponding simultaneous recording of the FM signal: the bottom is a zoomed FM signal with the 2 additional Cs temperatures 180°C and 190°C. The experiment is performed in a Cs cell ended by a sapphire window, whose temperature is ~300°C (see [1]), and the frequency markers are provided by the corresponding auxiliary SA reference.

Fig. 5: Predictions for $I_{fluo}$ from eq.(1) for various values of $\Lambda/\tau$ –as indicated—. The light dotted (red) line assumes no quenching at all (or $\Lambda/\tau \to \infty$), which is the standard situation for a (Doppler-broadened) Gaussian excitation lineshape, while the light (black) dashed line is for a quenching affecting symmetrically $v$ and $-v$ atoms, in analogy with a thin cell—see text. Except in this last case, the excitation lineshape is unaffected on the blue side by the $\Lambda/\tau$ value. The numerical calculation assumes $\lambda = 1$ μm, u = 200 m/s, $\tau = 100$ ns, $\gamma = 10$ MHz (*i.e.* $\gamma \ll ku$), yielding respectively for the absorption length $\Lambda$: 50 μm (red, dash-dotted), 20 μm (blue, dotted), 5μm (black, full line), 2 μm (green, dotted).

Fig.6: Predictions for $I_{fluo}$ assuming in eq. (1) that the absorption, maximal on-resonance ($\delta = 0$), decreases with the frequency detuning $\delta$ according to a Voigt profile —see on the right part of the figure the variations of the absorption length $\Lambda$ with the detuning. The spatial integration in eq. (1) is limited to a depth of observation of 3 mm. The numerical values are $\lambda = 1$ μm, u = 200 m/s (leading to ku = 200 MHz), $\tau = 30$ ns and $\Lambda(\delta = 0)$ as indicated. One has chosen $\gamma = 10$ MHz for all thick lines, while the light (blue) dash-dotted line is for $\gamma = 20$ MHz and $\Lambda(\delta = 0) = 20$ μm, showing that $\gamma$ ($\ll ku$) plays a role only when the effective absorption length $\Lambda(\delta)$ becomes comparable to or larger than the observation depth. The optical thicknesses as governed by $\Lambda(\delta = 0)$ values are chosen for resembling the single path absorption on the $D_1$ line of Cs respectively at 120°C, 130°C, 150°C —see fig.3.

Fig. 7: Same as fig.6 for $dI_{fluo}/d\delta$, which is the predicted FM signal for the excitation spectrum of fluorescence.

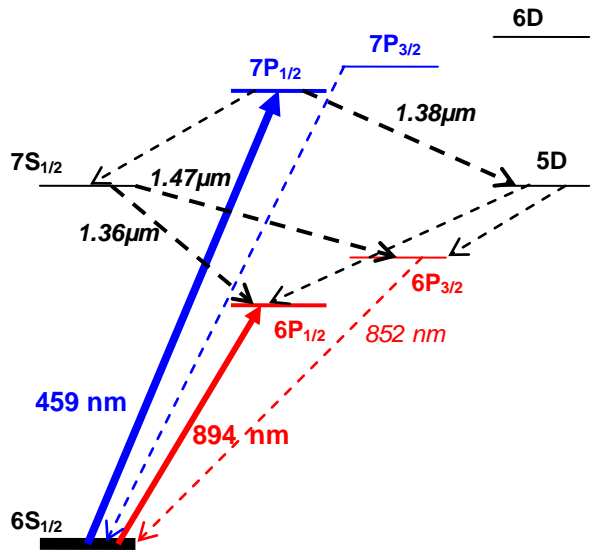

Figure 1

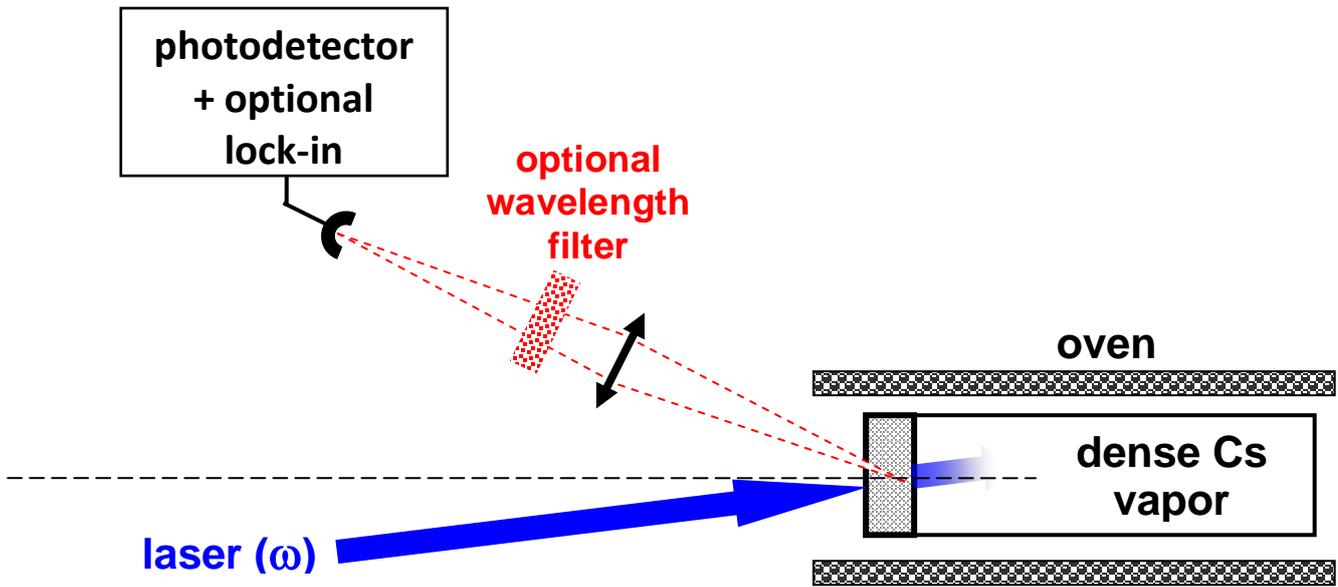

Figure 2

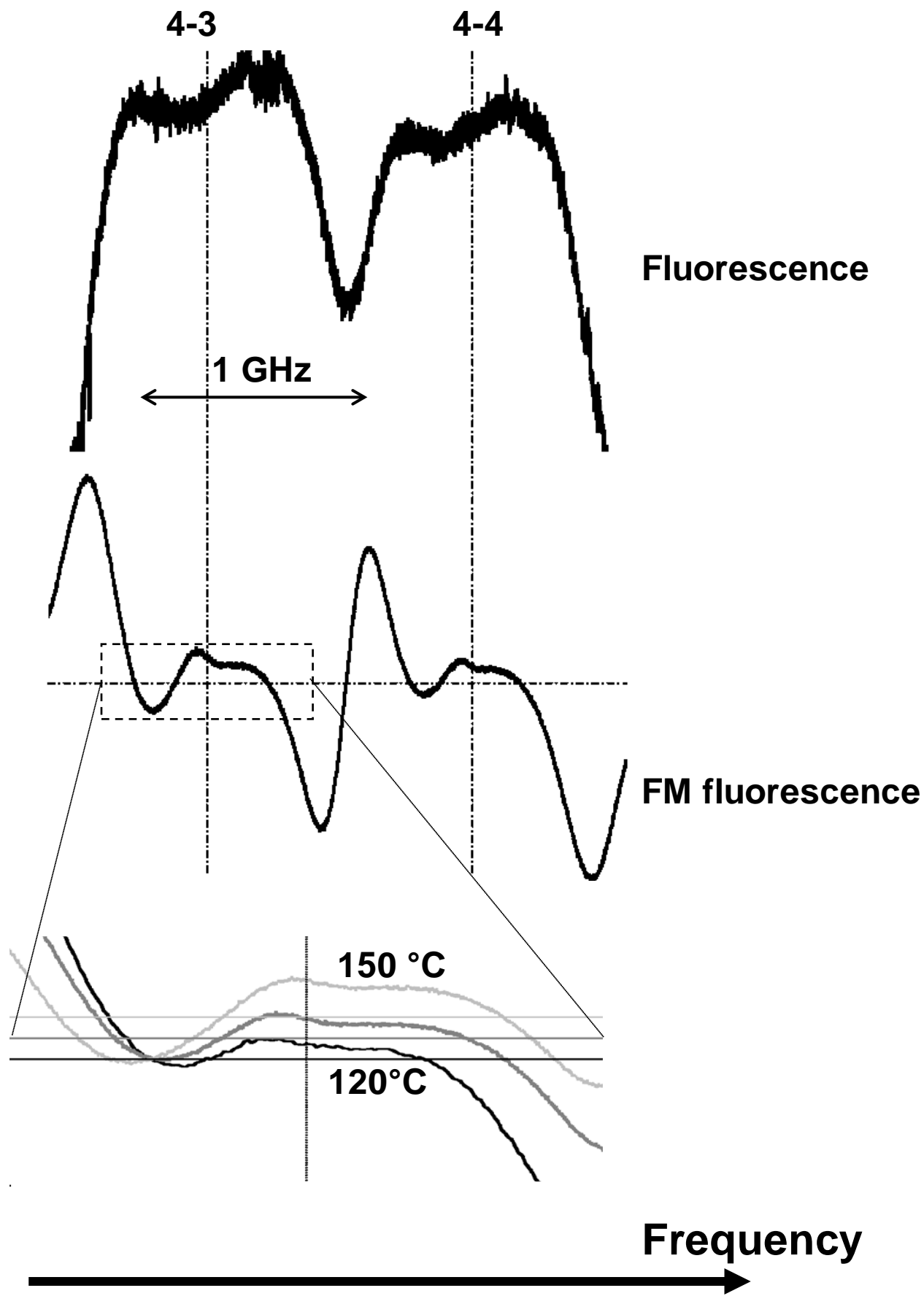

Figure 3

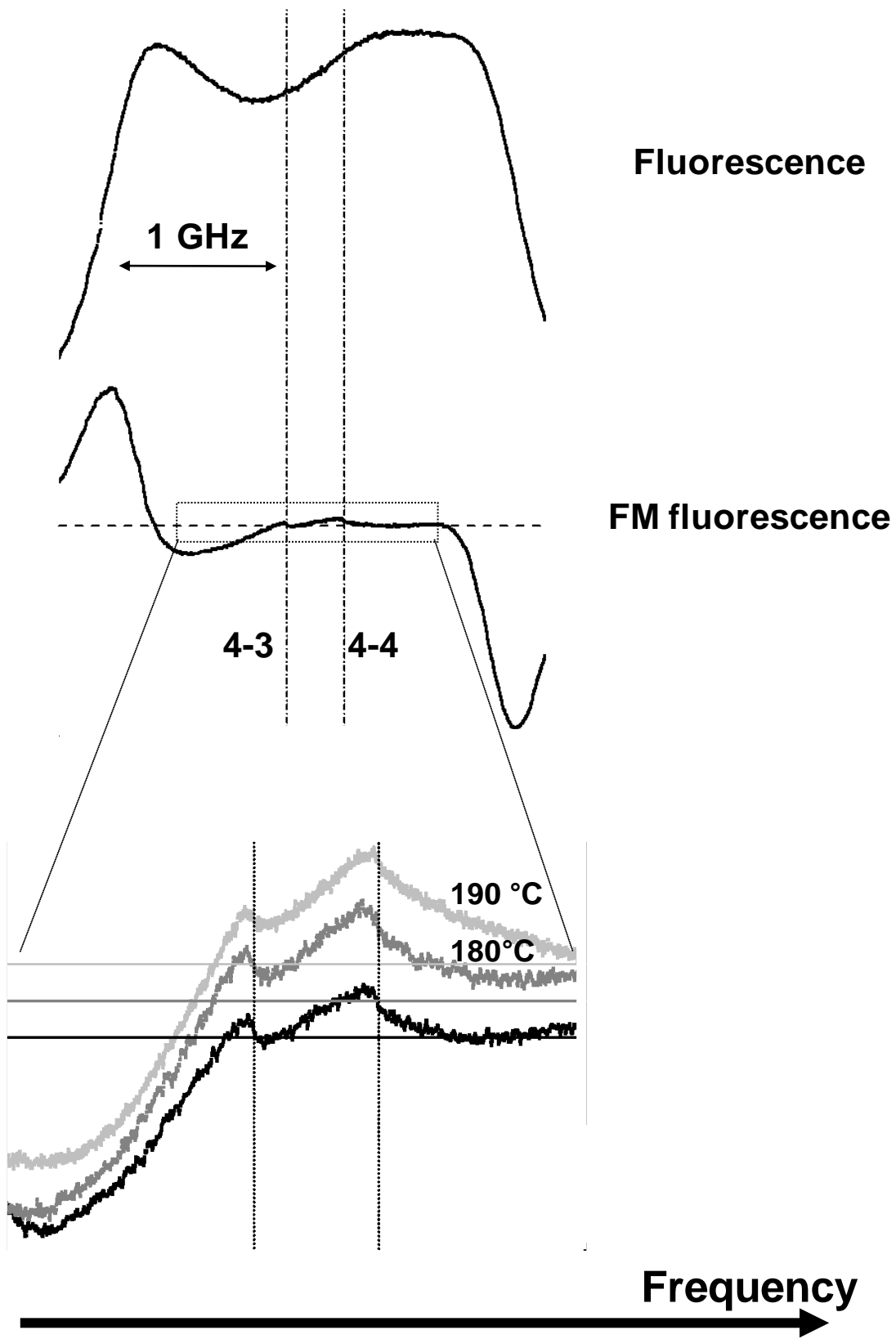

Figure 4

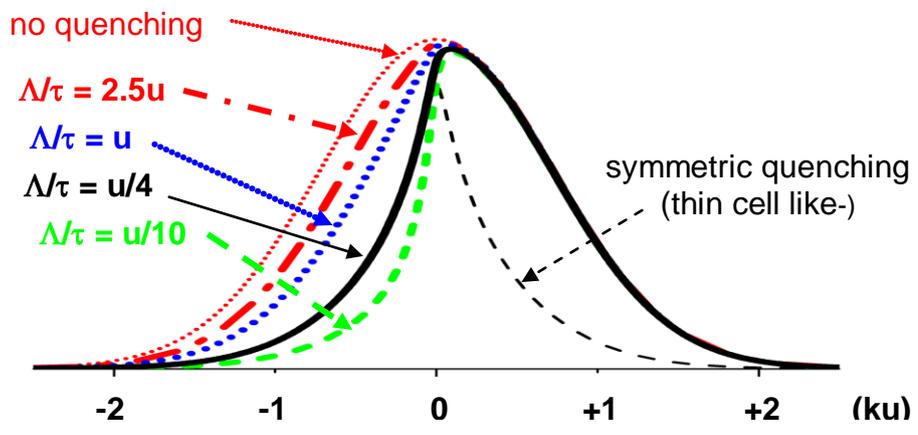

Figure 5

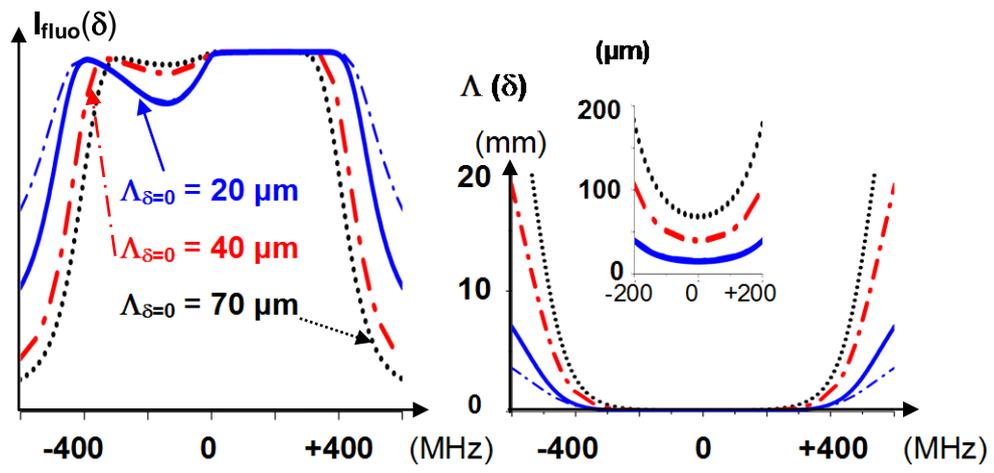

Figure 6

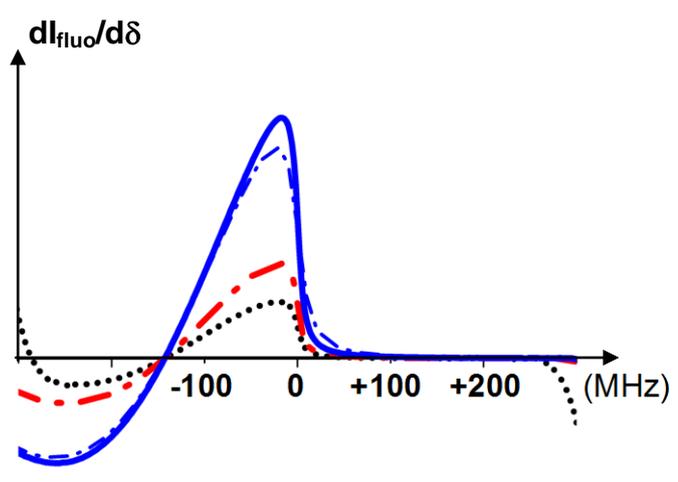

Figure 7